\documentclass[pra,twocolumn,showpacs,superscriptaddress,amsmath]{revtex4}

\usepackage{bm}
\usepackage{graphicx}

\newcommand{\bra}[1]{\langle #1 | \,}
\newcommand{\ket}[1]{\, | #1 \rangle}

\newcommand{\expv}[1]{\langle #1 \rangle}

\newcommand{\la}{\lambda}
\newcommand{\om}{\omega}
\newcommand{\Om}{\Omega}
\newcommand{\ga}{\gamma}

\newcommand{\de}{\delta}
\newcommand{\De}{\Delta}
\newcommand{\ka}{\kappa}
\newcommand{\eps}{\epsilon}
\newcommand{\veps}{\varepsilon}
\newcommand{\br}{\mathbf{r}}
\newcommand{\Eh}{\hat{\cal E}}
\newcommand{\Psih}{\hat{\Psi}}
\newcommand{\Sh}{\hat{S}}
\newcommand{\Ih}{\hat{\cal I}}
\newcommand{\sih}{\hat{\sigma}}
\newcommand{\zp}{z^{\prime}}
\newcommand{\tp}{t^{\prime}}

\begin{document}

\title{Strongly interacting photons in hollow-core waveguides}

\author{Ephraim Shahmoon}
\author{Gershon Kurizki}
\affiliation{Department of Chemical Physics, 
Weizmann Institute of Science, Rehovot 76100, Israel}
\author{Michael Fleischhauer}
\affiliation{Department of Physics and Research Center OPTIMAS,
Technische Universit\"at Kaiserslautern, D-67663 Kaiserslautern, Germany}
\author{David Petrosyan}
\affiliation{Department of Physics and Research Center OPTIMAS,
Technische Universit\"at Kaiserslautern, D-67663 Kaiserslautern, Germany}
\affiliation{Institute of Electronic Structure \& Laser, 
FORTH, GR-71110 Heraklion, Crete, Greece}

\date{\today}

\begin{abstract}
Hollow-core photonic-crystal waveguides filled with cold atoms can 
support giant optical nonlinearities through nondispersive propagation
of light tightly confined in the transverse direction. Here we explore 
electromagnetically induced transparency is such structures, considering
a pair of counter-propagating weak quantum fields in the medium of 
coherently driven atoms in the ladder configuration. Strong dipole--dipole
interactions between optically excited, polarized Rydberg states of the atoms
translate into a large dispersive interaction between the two fields.
This can be used to attain a spatially-homogeneous conditional phase shift 
of $\pi$ for two single-photon pulses, realizing a deterministic photonic 
phase gate, or to implement a quantum nondemolition measurement of the 
photon number in the signal pulse by a coherent probe, thereby achieving 
a heralded source of single or few photon pulses. 
\end{abstract}


\maketitle


Photons are ideal carriers of information in terms of transfer 
rates and distances. Yet, scalable and efficient quantum information 
processing \cite{QCcomp} with photons would require implementing 
deterministic quantum logic between single-photon qubits \cite{dpetr},
which is hindered by the weakness of optical nonlinearities in 
conventional media. Highly enhanced nonlinear interactions in atomic 
vapors \cite{imam} in the regime of electromagnetically induced transparency
(EIT) \cite{fllk,lukin,EIT} have emerged as a promising route to circumvent
these difficulties and to achieve large conditional phase shifts $\phi$ 
for pairs of slowly propagating photons. Attaining the phase shift of 
$\phi = \pi$ would amount to realizing the universal \textsc{cphase}
gate for photonic qubits \cite{QCcomp}.

Among the many relevant proposals 
\cite{lukimam,masalas,tripod,sanders,pbgeit,ddeit,gorshkov},
one of the most promising schemes is based on employing EIT in a ladder 
configuration \cite{ddeit}, wherein the photon-photon interaction is 
mediated by strong dipole--dipole interactions (DDIs) between 
optically excited Rydberg states of the atoms \cite{adams,rydrev}. 
An important advantage of this scheme is that the long-range nature 
of the DDI relaxes the need for tight focusing of the quantum fields 
to the atomic absorption cross-section $\varsigma \sim \la^2$,
which is close to the diffraction limit.

In Ref.~\cite{ddeit} we have presented an effective one-dimensional 
(1D) treatment of the dynamics of two slowly counter-propagating, 
weakly--focused single-photon pulses. We have done so by considering 
the electric fields only on the propagation axis, and have shown that, 
for a pair of photons passing through each other, the accumulated 
conditional phase shift $\phi$ can be both large and uniform in the 
longitudinal direction. In free space, however, the 1D treatment of 
interacting quantum fields is incomplete as it does not capture the 
diffraction effects and the fact that, in the transverse direction,
the resulting phase shift is inhomogeneous due to the relative coordinate 
dependence of the DDI potential \cite{simon}. To remedy these problems 
and achieve non-diffracting, uniform transverse phase-fronts, here 
we propose to impose onto the quantum fields only a single transverse 
mode by confining them into a hollow-core photonic-crystal waveguide 
\cite{knight,gaeta} filled with an ensemble of cold alkali atoms 
\cite{wgeitcold}. In what follows, we present a rigorous derivation
of 1D propagation equations for two interacting quantum fields.
We extend our earlier scheme by considering the atomic level
configuration involving different Rydberg states. We then discuss
the conditional phase shift for two single-photon pulses. Furthermore, 
we analyze a quantum nondemolition measurement (QND) of the photon 
number in the signal pulse inducing a phase shift of the coherent probe 
pulse. This can serve as a heralded source of single or few photon pulses.

\begin{figure}[t]
\includegraphics[width=8.5cm]{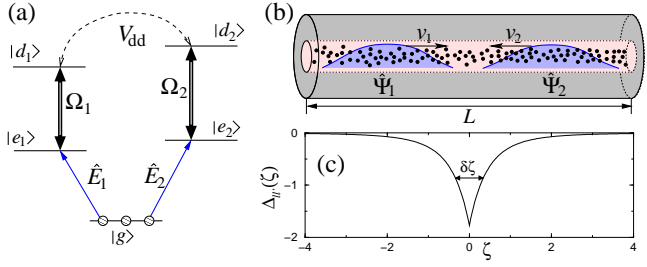}
\caption{(a)~Level scheme of atoms resonantly interacting with 
quantum fields $\hat{E}_{1,2}$ and classical driving fields $\Om_{1,2}$
on the corresponding transitions. $V_{\rm dd}$ denotes the DDI between 
atoms in Rydberg states $\ket{d}$.
(b)~The quantum fields transversely confined in a hollow-core 
waveguide of length $L$ filled with the atoms, counterpropagate 
as dark-state polaritons $\Psih_{1,2}$ having slow group velocities 
$v_{1,2}$ and interacting via long-range potential $\De_{12}(z_1-z_2)$ 
mediated by $V_{\rm dd}$.
(c)~The potential $\De_{ll^{\prime}}(\zeta)$ of Eq.~(\ref{1Dddpot}), 
as a function of dimensionless distance $\zeta$, in units of 
$2 C_{ll^{\prime}}/\hbar (\sqrt{2} w)^3$ Hz.} 
\label{fig:als}
\end{figure}

We begin by assuming that the transverse intensity profile of the 
counter-propagating fields $\hat{E}_{1}$ and $\hat{E}_{2}$ in the 
cylindrically symmetric waveguide is described by a Gaussian 
$e^{-r_{\bot}^2/w_f^2}$ of width $w_f$, where $r_{\bot} = |\br_{\bot}|$ 
is the distance from the propagation $z$ axis. The corresponding 
electric field can then be expressed as 
$\hat{E}_l(\br) = \veps_l e^{-r_{\bot}^2/2w_f^2} \Eh_l(z)$ ($l=1,2$), 
where $\veps_l = \sqrt{\hbar \om_l/2 \eps_0 V}$ is the field per photon
of frequency $\om_l$ within the quantization volume $V = \pi w_f^2 L$, 
with $L$ the waveguide length, while $\Eh_l(z) = \sum_k a_l^k e^{ikz}$
is the traveling-wave field operator, given by a superposition of 
bosonic operators $a_l^k$ for the longitudinal field modes $k$,
yielding the commutation relations 
$[\Eh_l(z),\Eh^{\dagger}_{l^{\prime}}(z^{\prime})] 
= L \de_{l l^{\prime}} \de(z - z^{\prime})$.  
An ensemble of $N$ cold atoms is trapped in the hollow core of the
waveguide \cite{wgeitcold}; the corresponding atomic density is then 
$\rho(\br) = (\pi w_a^2)^{-1} e^{-r_{\bot}^2/w_a^2} (N/L)$, where 
$w_a \, (\lesssim w_f)$ is the width of the transverse Gaussian distribution. 
The level configuration of the atoms, all of which are initially prepared in 
the ground state $\ket{g}$, is schematically shown in Fig.~\ref{fig:als}(a).
The quantum fields $\hat{E}_{1,2}$ resonantly interact with
the atoms on the transitions $\ket{g} \to \ket{e_{1,2}}$, respectively. 
The intermediate states $\ket{e_{1,2}}$ are resonantly coupled by two 
strong (classical) driving fields with Rabi frequencies $\Omega_{1,2}$ 
to the Rydberg states $\ket{d_{1,2}}$. In a static electric
field $E_{\rm st} \mathbf{e}_z$, these Rydberg states possess 
permanent dipole moments $\mathbf{p}= \frac{3}{2} n q e a_0 \mathbf{e}_z$,
where $n$ and $q$ are the (effective) principal and parabolic quantum 
numbers, $e$ is the electron charge, and $a_0$ is the Bohr radius 
\cite{RydAtoms}. A pair of atoms at positions $\br$ and $\br^{\prime}$ 
excited to states $\ket{d_l}$ and $\ket{d_{l^{\prime}}}$ interact with each 
other via a DDI potential $V_{\rm dd}$ resulting in an energy shift
\begin{equation}
\hbar \De_{ll^{\prime}}(\br -\br^{\prime}) = C_{ll^{\prime}} \, 
\frac{1 - 3 \cos^2 \vartheta}{|\br -\br^{\prime}|^3} ,
\end{equation}
where $\vartheta$ is the angle between vectors $\mathbf{e}_z$ 
and $\br^{\prime} - \br$, and 
$C_{ll^{\prime}} = \wp_{d_l} \wp_{d_{l^{\prime}}}/(4 \pi \eps_0)$
is proportional to the product of atomic dipole moments 
$\wp_{d_l} = \bra{d_l} \mathbf{p} \ket{d_l}$. We assume that state 
mixing within the same $n$ manifold is suppressed by a proper choice 
of parabolic $q$ and magnetic $m$ quantum numbers \cite{RydAtoms}.

We use collective atomic transition operators
$\sih_{\mu \nu}(\br) = 1/ N_{\br} \sum_{j=1}^{N_{\br}} 
\ket{\mu}_{jj}\bra{\nu}$ averaged over the volume element $\De V$ containing 
$N_{\br} = \rho(\br) \, \De V \gg 1$ atoms around position $\br$.
In the frame rotating with the frequencies of the optical fields, the 
interaction Hamiltonian $H = V_{\rm af} + V_{\rm dd}$ contains the 
atom-field and DDI terms 
\begin{subequations}
\label{HamCont}
\begin{eqnarray}
V_{\rm af} &=& - \hbar \! \int \! d^3 r \, \rho(\br) 
\sum_{l=1,2} \big[ g_l e^{-r_{\bot}^2/2w_f^2} \Eh_l(z) \sih_{e_l g}(\br) 
\nonumber \\ && \qquad \qquad \qquad + 
\Om_l \sih_{d_l e_l}(\br) \big] + \textrm{H. c.} ,  \\
V_{\rm dd} &=& \hbar \!  \int \! d^3 r \, \rho (\br) 
 \! \int \! d^3 r^{\prime} \rho (\br^{\prime}) 
\nonumber \\ && \;\; \times \frac{1}{2}  
\sum_{l,l^{\prime}=1,2} 
\sih_{d_l d_l}(\br) \De_{ll^{\prime}}(\br -\br^{\prime}) 
\sih_{d_{l^{\prime}} d_{l^{\prime}}} (\br^{\prime}) , \quad
\end{eqnarray}
\end{subequations}
where $g_l = (\wp_{ge_l}/\hbar) \veps_l$ is the corresponding atom-field 
coupling constant, with $\wp_{ge_l}$ being the dipole matrix element on 
the transition $\ket{g} \to \ket{e_l}$.

Using Hamiltonian $H$, we derive the Heisenberg-Langevin equations 
for the atomic operators $\sih_{ge_l}(\br)$, $\sih_{gd_l}(\br)$ and 
the propagation equations for the slowly-varying quantum fields $\Eh_l(z)$.
Solving for the atomic operators perturbatively in the small parameters 
$g_l \Eh_l/\Om_l$ and in the adiabatic approximation \cite{EIT,ddeit,fllk}, 
and after substituting into the equations for the fields, we obtain
the following propagation equations for the dark-state polaritons 
$\Psih_l = \sqrt{c/v_l} \, \Eh_l$ \cite{fllk}, 
\begin{equation}
\left(\frac{\partial}{\partial t} \pm v_l\frac{\partial}{\partial z}\right) 
\Psih_l(z,t) = - i  \sin^2 \theta_l \Sh_l(z,t) \Psih_l(z,t) , \label{Plprop} 
\end{equation}
the sign ``$+$'' or ``$-$'' corresponding to $l = 1$ or $2$, respectively, 
$v_l = c \cos^2 \theta_l$ is the group velocity of the corresponding 
field in the EIT medium, and the mixing angles $\theta_l$ are 
defined through $\tan^2 \theta_l = (g_l^2 N/|\Om_l|^2) (w/w_a)^2$,
with $w = w_a w_f (w_a^2 + w_f^2)^{-1/2}$. Operators $\Sh_l(z,t)$ are 
responsible for the self- and cross-phase modulation between the fields,
\begin{eqnarray}
\Sh_l(z,t) &=& \frac{1}{L} \int_0^L \!\! d \zp 
\big[ \De_{ll}(z - \zp) \sin^2 \theta_l \Ih_l(\zp, t) 
\nonumber \\ & & \qquad \quad 
+ \De_{ll^{\prime}}(z - \zp) \sin^2 \theta_{l^{\prime}} \Ih_{l^{\prime}} (\zp, t) \big],
\end{eqnarray}
where $\Ih_l \equiv  \Psih_l^{\dagger} \Psih_l  
= (c/v_l) \Eh_l^{\dagger} \Eh_l $ are the polariton intensity 
(excitation number) operators in the EIT medium, which correspond to 
the photon number operators outside the medium ($v_l=c$) \cite{fllk},
while the effective one-dimensional DDI potentials $\De_{ll^{\prime}}(z - \zp)$
result from $\De_{ll^{\prime}}(\br -\br^{\prime})$ upon double integration 
over the transverse coordinates,
\begin{eqnarray}
\De_{ll^{\prime}}(z - \zp) &=& \frac{1}{(\pi w^2)^2} 
\! \int \!\! d^2 r_{\bot} \!\! \int \!\! d^2 r^{\prime}_{\bot}
e^{- (r^2_{\bot} + r^{\prime 2}_{\bot})/w^2} \De_{ll^{\prime}}(\br - \br^{\prime}) 
\nonumber \\
&=& \frac{2 C_{ll^{\prime}}}{\hbar (\sqrt{2} w)^3} \big[ 
2 |\zeta| - \sqrt{\pi} (1+ 2 \zeta^2) e^{\zeta^2} \textrm{erfc} (|\zeta|) 
\big] , \label{1Dddpot} \\ 
&& \quad \zeta \equiv (z - \zp)/\sqrt{2} w . \nonumber
\end{eqnarray}
As seen in Fig.~\ref{fig:als}(c), $\De_{ll^{\prime}}(\zeta)$ is sharply 
peaked around $\zeta = 0$ with the range (FWHM) of $\de \zeta \simeq 0.65$. 

It follows from Eq.~(\ref{Plprop}) that the intensity operators $\Ih_l$ 
are constants of motion: $\Ih_l(z,t) = \Ih_l(z \mp v_l t,0)$, 
the upper (lower) sign corresponding to $l=1$ ($l=2$). The 
solution for the field operators then reads
\begin{equation}
\Psih_l(z,t) = \exp \bigg[- i \sin^2 \theta_l \!\! \int_0^t \!\!\! d \tp 
\Sh_l(z \mp v_l(t-\tp),\tp) \bigg] \Psih_l(z \mp v_l t,0) . \label{Psolv}
\end{equation}
The validity of this dissipation-free solution hinges on the following
assumptions: (i) The duration $T_l$ of each pulse exceeds the inverse 
of the corresponding EIT bandwidth 
$\de \om_l = |\Om_l|^2 / (\ga_{ge_l} \sqrt{\ka_{l} L})$,
where $\ga_{ge_l}$ is the transversal relaxation rate and 
$\ka_l \simeq \varsigma_l \bar{\rho}$ is the resonant absorption 
coefficient, with $\varsigma_l = 3 \la_l^2/(2 \pi)$ the absorption 
cross section on the transition $\ket{g} \to \ket{e_l}$ and 
$\bar{\rho} = N/[\pi(w_a^2+w_f^2)L]$ the effective atomic density. 
With $v_l = 2 |\Om_l|^2 /(\ka_{l} \ga_{ge_l})$, this yields the condition
$(\ka_l L)^{-1/2} \ll T_l v_l/L < 1$ which requires a medium with large 
optical depth $\ka_l L \gg 1$ \cite{lukin,EIT}. (ii) The DDI induced
frequency shifts lie within the EIT bandwidths, 
$\sin^2 \theta_l \expv{\Sh_l(z)} < \de \om_l, \, \forall \, z \in [0,L]$.
(iii) The propagation/interaction time of each pulse $t_{\mathrm{out}} = L/v_l$
is limited by the relaxation rate $\ga_{gd_l}$ of the $\sih_{gd_l}$ 
coherence via $t_{\mathrm{out}} \ga_{gd_l} \ll 1$. 

In what follows, we employ Eq. (\ref{Psolv}) to demonstrate the 
quantum phase gate between two single-photon pulses $\hat{E}_{1,2}$,
and to realize a quantum nondemolition measurement of photon number 
in the signal pulse $\hat{E}_2$ by a coherent probe pulse $\hat{E}_1$.  
For simplicity of notation, we set $\theta_{1,2} = \theta$, 
i.e., $g_1^2 N/|\Om_1|^2 = g_2^2 N/|\Om_2|^2$.


We are concerned with the evolution of input state 
$\ket{\Phi_{\rm in}} = \ket{1_1} \ket{1_2}$
composed of two single-excitation wavepackets $\ket{1_l} 
= \big[\frac{1}{L} \int \! dz f_l(z) \Psih_l^{\dagger}(z) \big] \ket{0}$
whose spatial envelopes inside the medium 
$f_{l}(z) = \bra{0} \Psih_l(z,0)\ket{1_l}$ are
normalized as $\frac{1}{L} \int \! dz |f_l(z)|^2 =1$. 
With the operator solution (\ref{Psolv}), for the (equal-time) 
correlation amplitude or the ``two-photon wavefunction''  
$F_{12}(z_1,z_2,t) = \bra{0} \Psih_1(z_1,t) \Psih_2(z_2,t)\ket{\Phi_{\rm in}}$
\cite{lukimam,dpetr} we obtain 
\begin{eqnarray}
F_{12}(z_1,z_2,t) = f_1(z_1 - vt) f_2(z_2 + vt) \exp[i \phi_{12}(z_1,z_2,t)] , \\
\phi_{12} (z_1,z_2,t) = - \sin^4 \theta \! \int_0^t \!\! d \tp 
\De_{12} \big( z_1 - z_2 - 2 v (t - \tp) \big) . \;\; \label{phiphsh}
\end{eqnarray}
Hence, the two polaritons counterpropagate in a shape-preserving manner 
with group velocities $\pm v$. Since $\Ih_l \Psih_l\ket{1_l} = 0$, 
the self-interaction within each pulse is absent, while the cross-interaction
between the pulses results in the phase-shift (\ref{phiphsh}).
Assume that at $t=0$ the first pulse is centered at $z_1 =0$ 
and the second pulse at $z_2 = L$, while after the interaction, 
$t_{\mathrm{out}} = L/v$, the coordinates of the two pulses are $z_1 = L$ 
and $z_2 = 0$, respectively. The accumulated phase-shift is then 
$\phi_{12}(L,0,L/v)=-\sin^4 \theta/ v \int_0^L \!\! d \zp \De_{12}(2\zp -L)$.
To evaluate the integral, we replace the variable 
$(2 \zp -L)/\sqrt{2}w \to \zeta^{\prime}$ and extend the integration 
limits to $L/(\sqrt{2}w) \to \infty$, obtaining
\begin{equation}
\phi_{12} = \frac{C_{12} \sin^4 \theta}{\hbar w^2 v} , \label{phshft}
\end{equation}
which is \textit{spatially uniform} and the state of the system at 
$t_{\mathrm{out}}$ is $\ket{\Phi_{\rm out}} = e^{\phi_{12}} \ket{\Phi_{\rm in}}$.
Since for input states $\ket{m_1} \ket{n_2}$ ($m,n=0,1$)
there is no phase shift when $m + n <2$, the conditional two-photon
phase shift $\phi_{12} = \pi$ is equivalent to the \textsc{cphase} 
gate $\ket{\Phi_{\rm out}} = (-1)^{mn} \ket{m_1} \ket{n_2}$ \cite{QCcomp}.


We next consider the probe pulse in a multimode coherent
state $\ket{\alpha_1} \equiv \Pi_{k} \ket {\alpha_1^k}$, 
which is an eigenstate of the field operator $\Psih_1(z)$ with 
eigenvalue $\alpha_1(z) = \sum_{k} \alpha_1^{k} e^{i k z}$: 
$\Psih_1(z)\ket{\alpha_1} = \alpha_1(z)\ket{\alpha_1}$.
The signal pulse can be in any superposition or mixture of 
the $n$-photon number states $\ket{n_2} = \frac{1}{\sqrt{n!}} 
\big[\frac{1}{L} \int \! dz f_2(z) \Psih_2^{\dagger}(z) \big]^n \ket{0}$.
Given an input state $\ket{\Phi_{\rm in}} = \ket{\alpha_1} \ket{n_2}$,
and neglecting the self-interaction, for the expectation value 
of the probe field we have 
\begin{multline}
\expv{\Psih_1(z,t)} = \alpha_1(z - vt) \\ 
\quad \times \bra{n_2} \exp \bigg[- i \frac{\sin^4 \theta}{L} 
\!\! \int_0^t \!\! d \tp 
\!\! \int_0^L \!\! d \zp \De_{12}(z -\zp - v (t-\tp)) \\ 
\times \Ih_{2} (\zp + v \tp, 0) \bigg] \ket{n_2}.   
\end{multline}
As before, we assume that at $t=0$ the probe and signal pulses 
are centered, respectively, at $z=0$ and $z=L$. The output probe 
field at $t_{\mathrm{out}} = L/v$ and $z=L$ is then
\begin{multline}
\expv{\Psih_1(L,L/v)} = \alpha_1(0) \\
\times \bra{n_2} \exp \bigg[ - i \frac{\sin^4 \theta}{L} 
\!\! \int_0^{L/v} \!\!\!\!\! d \tp 
\!\! \int_0^L \!\! d \zp \De_{12}(\zp - v \tp) \\ 
\times \Ih_{2} (\zp + v \tp, 0) \bigg] \ket{n_2} . \label{expvE}
\end{multline}
Recall that the DDI potential $\De_{ll^{\prime}}(z)$ is sharply peaked around 
$z=0$ with the range $\de z \lesssim w \ll L$ [Fig.~\ref{fig:als}(c)], while
$\int_{-\infty}^{\infty} d z \De_{ll^{\prime}}(z) = - 2 C_{ll^{\prime}}/(\hbar w^2)$. 
On the other hand, in the EIT medium, 
$\bra{n_l} \Ih_{l} (z) \ket{n_l} = n |f_l(z)|^2$ 
are smooth pulses of length $T_l v \lesssim L$.
To evaluate the integral in the exponential of Eq. (\ref{expvE}),
we may therefore replace the DDI potential as 
$\De_{12}(z) \to - 2 C_{12}/(\hbar w^2) \, \de (z)$. We then 
obtain $\expv{\Psih_1(L,L/v)} = \alpha_1(0) \exp(i \phi_{12} n_2)$, with
$\phi_{12}$ given by Eq. (\ref{phshft}). This indicates that, at the
output from the medium [$\Psih_1(L+0) = \Eh_1(L+0)$], the coherent 
probe field has acquired a phase proportional to the number of 
photons $n_2$ in the signal field. This phase can be measured 
by, e.g., a single-port homodyne detection using another coherent 
field of the same amplitude $|\alpha_1|$. The average detector signal 
is then $s(n_2) = 4|\alpha_1|^2 \sin^2(\phi_{12} n_2/2)$ with the 
corresponding uncertainty $\de s(n_2) = \sqrt{2 s(n_2)}$. 
Our aim is to distinguish with high probability the photon number 
states with $n_2 \in [0,n_{\textrm{max}}]$. This requires that 
$\phi_{12} n_{\textrm{max}} \leq \pi$, while the measurement 
uncertainty constraint yields 
$s(n_2) - s(n_2-1) > \frac{1}{2}[\de s(n_2) + \de s(n_2 - 1)]$.  

Above we have neglected the self interaction within the probe pulse,
which would otherwise dephase the coherent state. We can estimate
its effect as follows: 
$- \sin^4 \theta/L \int_0^{L/v} \!\! d \tp \!\! \int_0^L \!\! d \zp
\De_{11}(\zp - v \tp) |\alpha_{1} (\zp - v \tp)|^2 
\simeq 2 C_{11}\sin^4 \theta/(\hbar w^2 v) \, |\alpha_1(0)|^2$, 
which should be small compared to $\phi_{12}$. This leads 
to the condition $2 C_{11}|\alpha_1(0)|^2 < C_{12}$. 
(Note that, as long as we are concerned with determining the photon 
number in the signal field, its self-interaction is immaterial.)
Thus, for the QND measurement of the signal photon number by a 
coherent probe, the Rydberg states $\ket{d_{1,2}}$ should be chosen 
such that $\wp_{d_1} < \wp_{d_2}/(2 |\alpha_1(0)|^2)$, and therefore 
the self-interaction of the probe, $C_{11} \propto \wp_{d_1}^2$, is 
small compared to the cross interaction $C_{12} \propto \wp_{d_1}\wp_{d_2}$. 
On the other hand, to realize the \textsc{cphase} gate between two 
single photon pulses, $\phi_{12} = \pi$, both states $\ket{d_{1,2}}$ 
should have large and comparable dipole moments $\wp_{d_{1,2}}$ 
so that $C_{12}$ is large. 

Of course, in all cases we need to satisfy condition (ii), since otherwise
the DDI frequency shifts beyond the EIT transparency window would induce 
strong self and/or cross absorption of the fields \cite{wgeitcold}.
We therefore require that
\begin{equation} 
\frac{2 C_{ll^{\prime}} \sin^4 \theta}{\hbar w^2 L}  
\max \expv{\Ih_{l^{\prime}} (z)} < \de \om_{l} \quad (l,l^{\prime} = 1,2) . 
\label{condii}
\end{equation} 
In terms of the phase shift per photon $\phi_{ll^{\prime}}$, Eq.~(\ref{phshft}),
and assuming smooth $n_l$-photon pulses of lengths $T_l v \lesssim L$, we 
then have $2 \phi_{ll^{\prime}} \, n_{l^{\prime}} < T_{l^{\prime}} \de \om_l$ for 
cross-interaction and $2 \phi_{ll} (n_{l}-1) < T_{l} \de \om_{l}$ for self 
interaction. In turn, the product of the pulse duration and 
EIT bandwidth is restricted by the optical depth as
$T_{l^{\prime}} \de \om_{l} \lesssim \frac{1}{2}\sqrt{\ka_{l} L}$. 
We thus obtain that the maximal cross and self phase shifts are limited by 
\begin{equation} 
\phi_{ll^{\prime}} \, n_{l^{\prime}}, \, \phi_{ll} (n_{l}-1) 
< \frac{\sqrt{\ka_{l} L}}{4} .  
\end{equation}
Alternatively, the photon number in each pulse is limited by
\begin{equation}
n_l < \frac{\sqrt{\ka_{l^{\prime}} L}}{4 \phi_{ll^{\prime}} } , \;
\frac{\sqrt{\ka_{l} L}}{4 \phi_{ll}} + 1 . 
\end{equation}


To relate the foregoing discussion to a realistic experiment, we 
assume a hollow-core waveguide of length $L \sim 1\:$cm with the lowest
transverse mode of width $w_f \simeq 2\mu$m \cite{knight,gaeta,wgeitcold}. 
The waveguide is filled with $N \simeq 5\times 10^4$ cold Rb atoms 
tightly confined by a guided dipole trap to $w_a \simeq 2\mu$m, leading 
to the effective density $\bar{\rho} \simeq 2 \times 10^{11}\:$cm$^{-3}$.
For the two quantum fields tuned to the D1 and D2 transitions 
$\ket{g} \to \ket{e_{1,2}}$ ($\la_1 = 795\:$nm, $\la_2 = 780\:$nm),
the corresponding optical depths are $\ka_1 L \simeq 600$ and
$\ka_2 L \simeq 580$. With $\ga_{ge_1} \simeq 1.8 \times 10^7\:$s$^{-1}$,
$\ga_{ge_2} \simeq 1.9 \times 10^7\:$s$^{-1}$, and 
taking $\Om_{1} \simeq 7.35 \times 10^6\:$rad/s, 
$\Om_{2} \simeq 7.43 \times 10^6\:$rad/s, 
the group velocities are $v_{1,2} = 100\:$m/s.
The bandwidth of the pulses $T_l^{-1} \gtrsim v/L = 10^{4}\:$s$^{-1}$ is 
smaller than the EIT bandwidth $\de \om_l \simeq 1.2 \times 10^5\:$rad/s.
To realize the \textsc{cphase} gate, we choose the Rydberg states 
$\ket{d_{1,2}}$ with $\wp_{d_1} = \wp_{d_2} = 315 ea_0$ (quantum 
numbers $n=15$ and $q = n-1$), leading to the conditional phase shift 
$\phi_{12} = \pi$. For the QND measurement of photon number $n_2 \leq 2$ 
in the signal field with a weak coherent probe $|\alpha_1|^2 \simeq 4$,
the corresponding dipole moments for the Rydberg states are 
$\wp_{d_1} = 50 ea_0$ and $\wp_{d_2} = 450 ea_0$, leading to 
the cross-phase shift per photon of $\phi_{12} = 0.7$. 
We have verified that in both cases the DDI frequency shifts 
are within the EIT window $\de \om_l$ [cf. Eq.~(\ref{condii})]. 


Hence, the present scheme enables a realization of deterministic 
quantum gates with photonic qubits and is capable to distinguish 
with high probability the photon number states via QND measurement, 
which can serve as a heralded source of single of few photon pulses. 
In closing, we note that all the necessary ingredients of our 
proposal, including EIT via Rydberg states \cite{adams,rydrev} 
and in hollow-core waveguides \cite{knight,gaeta,wgeitcold}, 
have already been demonstrated experimentally.


\end{document}